\pdfoutput=1
\documentclass[conference]{IEEEtran}
\IEEEoverridecommandlockouts
\usepackage{subcaption}
\usepackage{cite}
\usepackage{amsmath,amssymb,amsfonts}
\usepackage{algorithmic}
\usepackage{graphicx}
\usepackage{textcomp}
\usepackage{xcolor}

\def\BibTeX{{\rm B\kern-.05em{\sc i\kern-.025em b}\kern-.08em
    T\kern-.1667em\lower.7ex\hbox{E}\kern-.125emX}}
\begin{document}

\title{Control Hardware-in-the-loop for Voltage Controlled Inverters with Unbalanced and Non-linear Loads in Stand-alone Photovoltaic (PV) Islanded Microgrids\\
%{\footnotesize \textsuperscript{*}Note: Sub-titles are not captured in Xplore and
%should not be used}
%\thanks{Identify applicable funding agency here. If none, delete this.}
}
%\author{\IEEEauthorblockN{Mehmet Emin Akdogan,  Mohammad Khatibi, Sara Ahmed}
\author{\IEEEauthorblockN{Mehmet Emin Akdogan, Sara Ahmed}
%\author{\IEEEauthorblockN{1\textsuperscript{st} Mehmet Emin Akdogan}
\IEEEauthorblockA{\textit{Electrical and Computer Engineering} \\
\textit{The University of Texas at San Antonio}\\
San Antonio, Texas \\
m.eminakdogan@gmail.com, sara.ahmed@utsa.edu}
%m.eminakdogan@gmail.com, mohammad.khatibi@utsa.edu, sara.ahmed@utsa.edu}
}
% \author{\IEEEauthorblockN{Mehmet Emin Akdogan}
% %\author{\IEEEauthorblockN{1\textsuperscript{st} Mehmet Emin Akdogan}
% \IEEEauthorblockA{\textit{Electrical and Computer Engineering} \\
% \textit{The University of Texas at San Antonio}\\
% San Antonio, Texas \\
% m.eminakdogan@gmail.com}
% \and
% \IEEEauthorblockN{ Mohammad Khatibi}
% \IEEEauthorblockA{\textit{Electrical and Computer Engineering} \\
% \textit{The University of Texas at San Antonio}\\
% San Antonio, Texas \\
% mohammad.khatibi@utsa.edu}
% \and
% \IEEEauthorblockN{ Sara Ahmed}
% \IEEEauthorblockA{\textit{Electrical and Computer Engineering} \\
% \textit{The University of Texas at San Antonio}\\
% San Antonio, Texas \\
% sara.ahmed@utsa.edu}
% }

\maketitle
\begin{abstract}
Unbalanced and nonlinear loads connected to micro grids (MG) with local distributed energy resources (DERs) are two of the leading causes of power quality problems. Nonlinear loads introduce voltage and current harmonics, and single phase loads can cause voltage and current imbalances in a three phase network. This paper presents a hierarchical control scheme for voltage controlled photovoltaic (PV) inverters with unbalanced and nonlinear loads in micro-grids.  The hierarchical control consists of primary control, voltage compensation control (VCC) and a DC voltage regulator (VR). The primary control scheme controls active and reactive power sharing and the VCC regulates the unbalanced voltage and harmonics distortion. The effectiveness of the scheme is verified using Opal-RT real-time simulation and experimentally using control hardware-in-the-loop. The voltage distortion at point of common coupling (PCC) decreased from 6.38 percent to 1.91 percent after compensation, while the unbalanced and harmonic load are shared proportionally among the DG units. 
% In addition, DC voltage regulator (VR) controllers fix DC link voltage in power stage of the distributed generations (DG) unit. A voltage controlled inverter based VCC method combined operation of the DC VR controller is proposed in the PV islanded micro-grid.
\end{abstract}


\section{Introduction}
In recent years, significant amount of renewable distributed energy resources (DERs) has been recently integrated into both bulk power transmission and distribution power systems to improve the sustainability of electric power systems.  Unlike traditional grids, renewable distributed generations (DG) can be installed in every location and require little maintenance effort and also operate locally through power electronics interface converter \cite{01,ali}. The increasing penetration of these inverter-based inertia-less DERs is rapidly changing the dynamics of large-scale power systems and causing several challenges such as inverse power flow, voltage deviation, and voltage fluctuation. To reduce the impact of high intermittent DER penetration, a micro grid (MG) is proposed \cite{02,03,04,05,06}. MG is a local grid that integrates multiple parallel DG units, energy storage and backup generators to improve the reliability of power system operation\cite{02,said}. A MG can become isolated from the grid during faults, which is called islanding mode\cite{03,04}.

PV-DGs in microgrids provide a clean and cost-effective solution for remote areas with no access to the utility grid such as, rural areas, marine, avionics and automotive. However, due to the intermittency of the PV systems, energy storage units are usually added to fulfill the immediate need for additional power to keep the system stable. However, integrating more PV units with the appropriate control could eliminate these storage units \cite{05,06,shah}. 

The control structure for PV-DGs usually consists of a primary controller \cite{04,07,08,13,14}. This controller comprises of a power droop controller, virtual impedance and current and voltage controllers. Every DG unit should operate independently while in parallel without communication due to long distance between DGs. Therefore, the droop control approach has been implemented widely in order to control the active and reactive power among the inverters and generate inverter reference voltage and frequency for PV-DGs \cite{04,07,08}. However, droop control is unable to share reactive power and harmonic current properly, thus virtual impedance is implemented to enhance the operation of multiple parallel DGs units\cite{08,09,10,11}. 

When PV-DGs feed unbalanced and nonlinear loads, the loads generate unbalanced voltage and voltage harmonic distortion respectively. They cause over voltage, overheating and deterioration of power quality on electrical equipment. Those power quality problems on power systems need to be improved. However, primary controllers are not enough to compensate for the power quality problems. Multiple micro grid control approaches are adopted to DG interface inverters in the literatures \cite{02,06,13,14,15,16,17,18,09,20,21,22,24}. 

\begin{figure}[!t]
\centerline{\includegraphics[width=\linewidth]{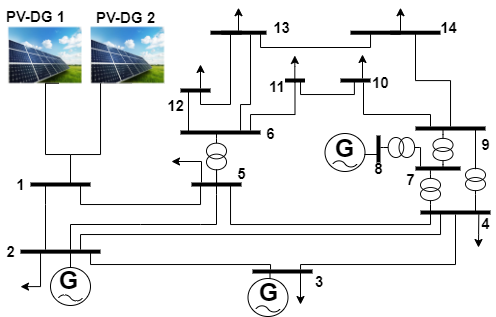}}
%\centering
%\includegraphics[scale=0.5]{pictures/14 bus network updated.png}
\caption{Single line diagram of the IEEE 14 node distribution system with parallel PV-DG units.}
%\hspace{\fill}
\vspace{-11pt}
\label{14bus}
\end{figure}

Some work has been proposed to reduce unbalanced voltage and harmonics distortion using active power filters (APF) in \cite{15,16}. APFs inject compensating harmonic current in opposite phase to cancel voltage harmonics of the APF installation point. However, the initial and operational costs of the power-electronic-based active power filter are very high.

Some control methods regulate power quality problems and reduce harmonic distortion for sensitive load bus (SLB) under unbalanced and nonlinear loads \cite{02,17,18,09,20,akdogan}. In \cite{02}, the authors discussed the negative and zero sequence current compensating controllers in islanded micro grids operating under unbalanced conditions but did not use negative sequences compensation of the PCC voltage to improve sharing of the unbalance voltage of the islanded micro grid. The approach presented in \cite{17} is based on applying a secondary control to provide better voltage quality.  Unbalanced harmonic compensation (UHC) is adopted to reduce unbalanced harmonic voltage at SLB. A hierarchical control scheme is proposed to improve power sharing and to perform voltage compensation of multi DERs micro grids including non-linear and unbalanced loads using radial basis function neural network-based harmonic power-flow calculations in \cite{18}. Power sharing enhancement control method in \cite{09} and also a fast harmonics suppression strategy in \cite{20} have been proposed to improve current sharing and to compensate reactive, unbalance and harmonic power sharing problems in an islanded ac micro grid. Although power quality at the selected bus can be improved, unbalance of PCC voltage is not compensated to generate compensation reference in the compensation effort controller.

There are two types of PV-DGs that can be classified into current source inverters (CSI) and voltage source inverter (VSI) \cite{05,07,amin}. CSIs are expected to inject maximum PV power into the grid or directly supply the loads. In \cite{21}, a voltage detection based harmonic compensator (HC) was proposed for CSIs based on the discrete Fourier transform (DFT) to better regulate the system harmonics. In \cite{22}, the authors presented a CSI under unbalanced and nonlinear loads in a grid connected PV-MG. However, a PV-DG in the islanded mode is not capable of providing AC voltage and frequency regulation through CSI \cite{06,23}. %Because there is no any grid system to support islanding PV-DG system. VSI should be considered for PV in the islanding mode to satisfy the voltage and frequency management, active and reactive power, and power quality in the system.
In addition, a harmonic load compensation scheme in \cite{24} is presented based on the voltage control method and a coordinated control of CSI and VSI units is proposed for reactive power sharing and voltage harmonics compensation in \cite{13}. These literature \cite{13,21,24} only considered harmonic compensation and not voltage unbalance. 

In addition, while PV inverters can behave as voltage sources to supply unbalanced harmonic loads in islanded mode to regulate inverter output voltage, DC link voltage cannot be controlled by the PV inverter \cite{03,Baranwa}. If the generated power from the PV modules is higher than the demanded power, a DC voltage regulator controller in \cite{03} is applied to a dc/dc converter to avoid an increase in the DC link voltage by curtailing PV output power. 

The impact of DC link voltage control combined with unbalanced distortion compensation based on voltage-source inverter has not been fully investigated in stand-alone systems. Therefore, this paper presents a complete control scheme combining voltage compensation controller with a DC voltage regulator controller under both unbalanced and nonlinear loads for voltage-controlled inverter based islanded PV systems with no storage units. The proposed hierarchical control includes primary controllers regulating inverter voltage and voltage compensating controller (VCC). While droop control regulates the active and reactive power of the PV DGs considering their rated power capacities, virtual impedance is considered to achieve better power sharing of reactive, unbalance and harmonic powers in the primary level. The VCC is proposed to compensate negative sequences of fundamental and main harmonics voltage at PCC. Also a DC voltage regulator controller is adopted in power stage of the DG unit.  The proposed control method is verified using Opal-RT real-time simulation of an IEEE-14 bus distribution system and experimentally using control hardware-in-the-loop. %Also a DC voltage regulator controller is adopted to fix Dc link voltage in power stage of the DG unit.

The remaining parts of the paper are presented as follows: in Section II, the system description and proposed hierarchical control scheme, including the primary controller and the VCC, are presented. Section III is dedicated to present the simulation and experimental results and finally, this paper is concluded in Section IV.
%Section III is dedicated to present the VCC scheme. 
%Simulation  and  experimental results  are  presented  in  Section  IV are presented in Section III.
\begin{figure*}[htbp]
\centering
\includegraphics[width=\linewidth]{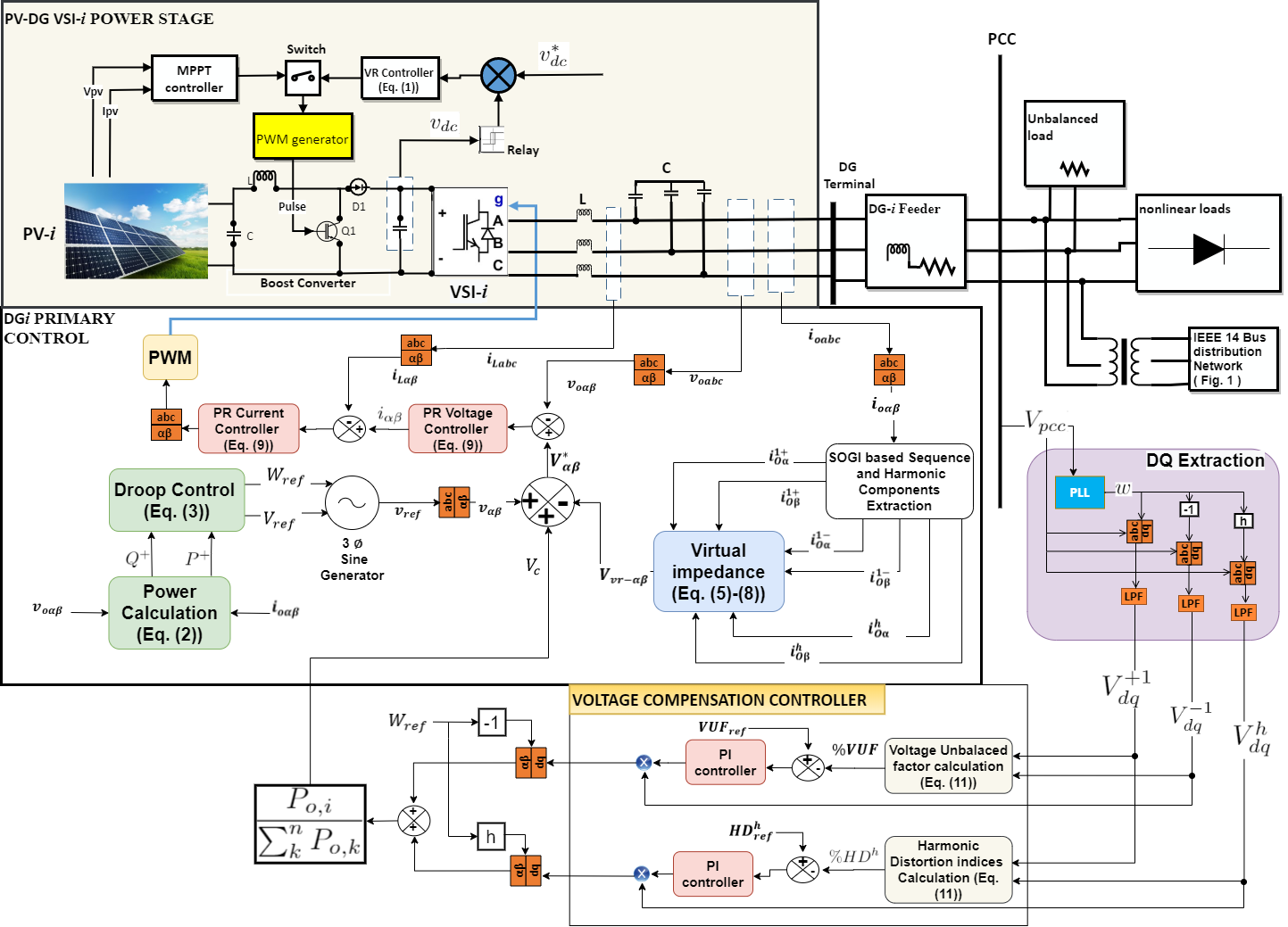}
\caption{Block diagram of hierarchical control scheme for one distributed generation in the proposed islanded Micro Grid.}
\label{main}
\vspace{-11pt}
\end{figure*}

\section{system description and proposed hierarchical control scheme}
Fig.~\ref{14bus} shows the single-line diagram of the system under study. The system consists of an IEEE-14 bus distribution network with two parallel PV-DG units with different power ratings and feeding unbalanced and nonlinear loads. %The proposed parallel PV-DG units is applied to the IEEE-14 bus distribution network to perform effect of the distribution network. %(The network consists of 3 small synchronous machines, 10 loads in the system totaling 140 MW at 12.47 kV) 
%you may need to move above paragraph to results  

The power stage of the PV-DG and the proposed hierarchical control scheme block diagram are demonstrated in Fig.~\ref{main}. The proposed hierarchical control consists of two control levels, the primary control level and the VCC level. The power stage of each DG unit in Fig.~\ref{main} includes PV solar panels connected to a boost dc/dc converter feeding a voltage source inverter with a LC filter. The primary control is responsible for power sharing between DGs. It comprises of power droop controller, virtual impedance, and proportional resonant (PR) controllers. The VCC is for PCC voltage quality improvement. The DGs feed multiple loads, for example, unbalanced and nonlinear loads. 
%There is a feeder between DG and loads to flow current smoothly. Moreover there are some complex load conditions such as balance/unbalanced and linear/nonlinear loads connected to PCC also. There are two voltage and current measurement points as DG terminal and PCC where DG terminal is output of filter and PCC is integrated point of the loads. DC link Voltage $(v_{dc})$ is provided by renewable energy source such as solar power and /or energy storage systems in order to take the small variations of dc link voltage $(v_{dc})$ in generation of the inverter gate signals by pulse width modulator (PWM).

% It is worth noting that the normal mode for triple harmonics is to be zero sequence. However, the unbalanced load will be added so   $3^{rd}$ harmonic compensation will be considered. In addition, the  $5^{th}, 7^{th}$ and $11^{th}$ harmonics (the main orders) of PCC voltage are considered. 
 %It should be noted that grid-connected mode is out of the scope of this paper since it only focuses on islanded mode.
%It should be noted that micro grid can be disconnected from utility grid during main grid faults.%
%In addition, each DG unit has a different power rating.
For facilitation, only DG1 details are shown in details in Fig.~\ref{main}. The detailed control loops will be discussed in this section. 

\subsection{PV DC/DC Converter Controller}
The boost dc/dc converter controller has two modes of operation, i.e. a maximum power point tracking (MPPT) mode and a DC voltage regulation (VR) mode (i.e, PI controller). 
\subsubsection{MPPT} In this mode, the controller uses incremental conductance technique to output maximum power of the PV array if the maximum PV power is equal to the demand power load. However, if the generated power from the PV modules is grater than the load demand, the PV output power has to be curtailed since the system is in islanded mode. This will prevent an increase in the DC link voltage $(v_{dc})$. In this case, the VR controller is activated to control the DC link voltage. 
\subsubsection{VR} The VR is activated automatically, when the $v_{dc}$ exceeds the reference of the DC link voltage ($v_{dc}^{*}$)\cite{02,05}.
%the DC-DC converter control is able to automatically switch from the MPPT controller to the VR controller \cite{02,05}. In other words, the output power of the PV array is bounded between the maximum and curtailed power in the islanded mode. Finally,
The VR controller generates a reference duty cycle ($D$) as shown in \eqref{VDC} where $k_{pdc}$ and $k_{idc}$ are the control parameters of the proportional integral (PI) of the DC link voltage, respectively.
\begin{equation}
D=k_{pdc}(v_{dc}^{*}-v_{dc)})+k_{idc}\int (v_{dc}^{*}-v_{dc)}) {\mathrm{d} t}
\label{VDC}
\end{equation}
\subsection{Voltage Source Inverter Primary Controller}
The detailed structure of the DG inverter primary controller is depicted in Fig.~\ref{main}. The primary controller is responsible for power sharing between DGs, regulating the frequency and amplitude of the DG output voltage reference through the following: 
\subsubsection{Power Calculation}
The three phase instantaneous active power ($p$) and reactive power ($q$) are calculated using measured output voltage ($v_{o\alpha \beta}$) and output current ($i_{o\alpha \beta}$) in $\alpha \beta$ reference frame. The power equations are shown in (\ref{q}) as
\begin{equation}
\begin{split}
p=\frac{3}{2}(v_{o\alpha }.i_{o\alpha }+v_{o\beta}.i_{o\beta})
\\
q=\frac{3}{2}(v_{o\beta }.i_{o\alpha }-v_{o\alpha }.i_{o\beta}).
\end{split}
\label{q}
\end{equation}

$p$ and $q$ are passed through first order low pass filters (LPF) with 2 Hz cut-off frequency to extract fundamental positive sequence active  ($P^{+}$) and reactive ($Q^{+}$) power respectively \cite{04,09}. In other words,  $P^{+}$ and $Q^{+}$ are the average values of instantaneous active and reactive power.  

\subsubsection{Droop Control}
The droop control regulates the active and reactive power. In addition, it controls the power sharing between DGs based on each DG unit power rating and the total load demand without communication between DGs \cite{04,07,08}. 
%The droop control strategy regulates  is adopted to generate the inverter reference voltage and frequency by adjusting fundamental positive active and reactive powers in islanded micro grid.the droop control strategy controls DG units power sharing
%it controls each DG unit to be proportional to its power rating in the islanded mode to share the total load demand without communication between DGs and avoid DG overloads.
The droop control characteristics are defined as: 
\begin{equation}
\begin{split}
W_{ref}=W^{*}-(P^{+})m_{p}
\\
V_{ref}=V^{*}-(Q^{+})n_{p}
\end{split}
\label{Vref}
\end{equation}
where $V^{*}$ and $W^{*}$ represent the amplitude and frequency of the output phase voltage reference and $m_{p}$ and $n_{p}$ represent active and reactive power proportional coefficient respectively. In addition, the ratio of power sharing among DGs are given by:
\begin{equation}
\begin{split}
\frac{P_{1}^{+}}{P_{i}^{+}}=\frac{m_{p1}}{m_{pi}}
\\
\frac{Q_{1}^{+}}{Q_{i}^{+}}=\frac{n_{p1}}{n_{pi}}
\end{split}
\end{equation}
where the number of suffixes demonstrate the number of each DG. It should be noted that the DG units in the system under study have different capacities. DG2 unit is double the power capacity of DG1 so droop control parameters of DG1 are double that of DG2 parameters as shown in Table~\ref{table}. Afterwards, the reference of the droop control ($v_{ref}$) in Fig.~\ref{main} is generated and transformed to $\alpha \beta$ reference frame to inject for DG voltage controller.

\begin{table}[t]
\caption{Power stage and Control Parameters}
\vspace{-2pt}
\begin{center}
\label{table}
\begin{tabular}{|c|c|}
\hline
\textbf{System Parameter}        & \textbf{Value (DG1/DG2)} \\ \hline
Switching frequency, $v_{dc}^{*}$,$w_{f}$   &    10 kHz, 600 V, 370 rad/s \\ \hline
PV Power &   3000/6000 W \\ \hline
\textbf{Power Control Parameter}        & \textbf{Value (DG1/DG2)} \\ \hline
$m_{p}, n_{p}$                   &         12e-4/6e-4, 1e-/0.5e-3       \\ \hline
$V^{*}$, $W^{*}$                            &      120 rms V, 370 rad/s \\ \hline
$VUF_{ref}$ $HD_{ref}^{h}$     &  0.2\%, 0.2\%, \\ \hline
$R_{v}^{1+}$,$R_{v}^{1-}$,  $L_{v}^{1+}$   &  0.3, 0.4/0.15,0.2 $\Omega$ 0.5/0.25 H       \\ \hline
$R_{v}^{+3}$, $R_{v}^{-5}$, $R_{v}^{+7}$, $R_{v}^{-11}$ & 3, 1, 1, 0.5/0.15, 0.5, 0.5, 0.25 $\Omega$     \\ \hline
\textbf{VCC PI Control Parameter} & \textbf{Proportional/	Integral Value} \\ \hline
Fundamental negative sequence    &      0.1/1.5            \\\hline
$3^{rd}$ Harmonic positive sequence  &    0.2/2          \\ \hline
$5^{th}$ Harmonic  negative sequence &     5/30         \\ \hline
$7^{th}$ Harmonic positive sequence  &     5/25         \\\hline
$11^{th}$ Harmonic  negative sequence  &      0.1/1      \\\hline
\end{tabular}
\vspace{-11pt}
\end{center}
\end{table}
\subsubsection{Virtual Impedance Controller} The droop control is unable to share reactive power and harmonic current proportionally and therefore the virtual impedance method is implemented to improve the reactive power sharing among DGs and also to share unbalanced and harmonic load current among DGs.\cite{08,09,10,11}.

The second-order general integrator (SOGI) is designed to extract the positive and negative sequences of the fundamental and harmonic components of the distorted inverter current using phase locked loop (PLL) algorithm and the detail of the SOGI is presented in \cite{25,26}.

Virtual impedance comprises of three loops, virtual positive sequence impedance (VPI) that is designed to improve the real and reactive power sharing, virtual negative sequence impedance (VNI) implemented to reduce effectiveness of the fundamental negative sequence current among DGs and virtual variable harmonics impedance (VVHI) that is also used for sharing harmonic power properly among DG units\cite{09,17}.

\begin{figure*}[t]
\begin{subfigure}{.48\textwidth}
  \centering
  % include first image
  \includegraphics[width=\linewidth]{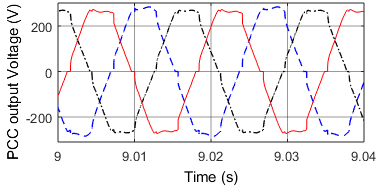} 
  \caption{}
  \label{Vpcca}
\end{subfigure}
\begin{subfigure}{.48\textwidth}
  \centering
  % include second image
  \includegraphics[width=\linewidth]{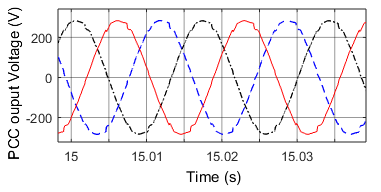}  
  \caption{}
  \label{Vpccb}
\end{subfigure}
\caption{Output Voltage of PCC before (a) and after (b) Proposed Method.}
\label{Vpcc}
\end{figure*}
\begin{figure*}[ht]
\begin{subfigure}{.4\textwidth}
  \centering
  % include first image
\includegraphics[width=\linewidth]{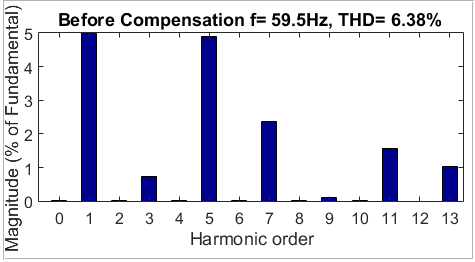} 
  \caption{}
  \label{thdb}
\end{subfigure}
\qquad
%\qquad 
\qquad \qquad\qquad  
\begin{subfigure}{0.4\textwidth}
  \centering
  % include second image
  \includegraphics[width=\linewidth]{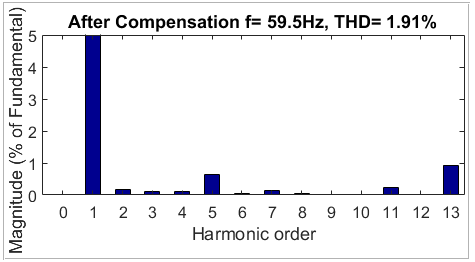}  
  \caption{}
  \label{thda}
\end{subfigure}
\caption{THD of PCC Voltage before (a) and after (b) the Compensation Method.}
\label{thd}
\vspace{-15pt}
\end{figure*}
%Note that only the  VPI loop has virtual inductance and resistance. 
VPI, VNI and VVHI loops in $\alpha\beta$ reference frame are derived respectively in ($\ref{VPI}$), ($\ref{VNI}$) and ($\ref{VVHI}$) 
\begin{equation}
\begin{split}
v_{v\alpha }^{1+}=R_{v}^{1+}.i_{o\alpha }^{1+}-L_{v}^{1+}.w_{f}.i_{o\beta }^{1+}
\\
v_{v\beta }^{1+}=R_{v}^{1+}.i_{o\beta}^{1+}+L_{v}^{1+}.w_{f}.i_{o\alpha}^{1+}
\end{split}
\label{VPI}
\end{equation}
\begin{equation}
\begin{split}
v_{v\alpha }^{1-}=R_{v}^{1-}.i_{o\alpha }^{1-}
\\
v_{v\beta }^{1-}=R_{v}^{1-}.i_{o\beta}^{1-}
\end{split}
\label{VNI}
\end{equation}
\begin{equation}
\begin{split}
v_{v\alpha,h }=R_{v}^{h}.i_{o\alpha }^{h}
\\
v_{v\beta,h }=R_{v}^{h}.i_{o\beta}^{h}
\end{split}
\label{VVHI}
\end{equation}
where $i_{o\alpha\beta }^{1\mp}$ represents the output current for fundamental positive and negative sequences in $\alpha\beta$ frame. Furthermore, $i_{o\alpha\beta}^{h}$ is  the $h^{th}$ harmonic sequence of output current in $\alpha\beta$ frame and h denotes dominant harmonic components. In addition, $R_{v}^{1+}$ and $L_{v}^{1+}$ are the virtual resistance and inductance for fundamental positive components, $R_{v}^{1-}$ represents the virtual resistance for fundamental negative components. Also $R_{v}^{h}$ is virtual resistance and for $h^{th}$ main harmonic components. Finally, $w_{f}$ represents the system fundamental frequencies \cite{09,17}. As shown in Fig.~\ref{main}, all three impedance loops added together to generate the voltage reference of  the virtual impedance $V_{vr-\alpha \beta }$ as follows:
\begin{equation}
v_{v\alpha,h }=v_{v\alpha }^{1+}+v_{v\beta }^{1+}+v_{v\alpha }^{1-}+v_{v\beta }^{1-}+v_{v\alpha,h }+v_{v\beta,h }
\label{Virtualref}
\end{equation}
where $v_{v\alpha\beta }^{1+}$  and $v_{v\alpha\beta }^{1-}$  are virtual voltage for fundamental positive and negative sequences in $\alpha\beta$ frame. Furthermore, $v_{v\alpha\beta,h }$ is the virtual $h^{th}$ harmonic sequence in $\alpha\beta$ frame.

\subsubsection{Proportional Resonant Controller} 
Current and voltage controllers are designed to regulate filter capacitor voltage and induct current in the stationary reference frame. Compared with PI control, PR voltage and current control can provide larger gain at the fundamental component %and the other frequencies 
and no phase shift which will help eliminate the steady-state error \cite{19}. PR voltage and current controllers are shown in (\ref{PR}) as
\begin{equation}
\begin{split}
G_{V} (s)= k_{pV}+\sum_{k=1,3,5,7}(\frac{2.k_{rVk}.w_{cV}.s}{s^{2}+2.w_{cV}.s+{(k.w_{o})}^{2}} )
\\
G_{I} (s)= k_{pI}+\sum_{k=1,3,5,7}(\frac{2.k_{rIk}.w_{cI}.s}{s^{2}+2.w_{cI}.s+{(k.w_{o})}^{2}} )
\end{split}
\label{PR}
\end{equation}
where $k_{pV}$ and $k_{pI}$ are the proportional coefficient of the voltage and current controller and $k_{rVk}$ and $k_{rIk}$ are the $k^{th}$ harmonic (including fundamental component as first harmonic) resonant coefficient of the voltage and current controller. $k_{pV}$ and $k_{pI}$ terms are chosen to meet the controller dynamics requirements, i.e., bandwidth, phase and gain margins. $w_{cV}$ and $w_{cI}$ are the cut-off frequency of the voltage and current controllers respectively. In addition, multiple harmonic compensation regulators at desired harmonic sequences ($3^{th}, 5^{th}, 7^{th}$) are added in parallel to the main PR controller to suppress voltage harmonics and to track harmonic current and voltage. 

The reference of the DG output voltage in $\alpha \beta$ frame ($V_{\alpha \beta }^{*}$) is provided by reference signals of the droop controller ($v_{\alpha \beta }$), the virtual impedance loop ($V_{vr-\alpha \beta }$) and the voltage compensation controller  compensation ($V_{c}$).  The reference of the DG output voltage is given by:
\begin{equation}
V_{\alpha \beta }^{*}=v_{\alpha \beta }-V_{vr-\alpha \beta }+V_{c}.
\label{Vo}
\end{equation}

Instantaneous output voltage ($v_{oabc}$) is measured and transformed to $\alpha \beta$ frame ($v_{o\alpha \beta}$). Then, $V_{\alpha \beta }^{*}$ is compared with $v_{o\alpha \beta}$. The error is received by PR voltage controller to generate the reference current ( $i_{\alpha \beta }$). The LC filter inductor current ($i_{Labc}$) is transformed to $\alpha \beta$ frame ($i_{L\alpha \beta }$) and is compared with $i_{\alpha \beta }$ to be controlled by the current controller. The output of the controller is transformed back to abc frame to produce three phase voltage reference for the pulse width modulator (PWM) block. Finally, the PWM controls the switching of the inverter based on this reference. 

\subsection{Voltage Compensating Controller (VCC)}
Nonlinear and unbalanced loads introduce harmonics and unbalance into the system voltage and current. The proposed VCC controller compensates for this voltage at PCC. Dq extraction block and the VCC are depicted in Fig.~\ref{main}.  

The load voltage ($V_{pcc}$) is extracted by the dq extraction block. Signs of “+”, “-” and “h” represent positive and negative sequence of fundamental component and $h^{th}$ harmonic component, respectively. For example, “$v_{dq}^{1-}$” is the negative sequence of the fundamental voltage in $dq$ frame. The PLL block is used to detect voltage frequency and angular frequency ($w$) of the system. $w$ is multiplied by positive and negative sequence of the fundamental component and selected $h^{th}$ harmonic component gain which represent 1, -1, +h, -h respectively. The park transformation is used to transform the $V_{pcc}$ from $abc$ to $dq$ frames. Then, second-order low-pass filter is applied with a cutoff frequency of 5 Hz and damping ratio of 2.5 to extract positive and negative sequences of the PCC voltage fundamental and main harmonic components in Fig.~\ref{main}.
\begin{figure*}[t]
\begin{subfigure}{.4\textwidth}
  % include first image
 \centering
 \includegraphics[width=\linewidth]{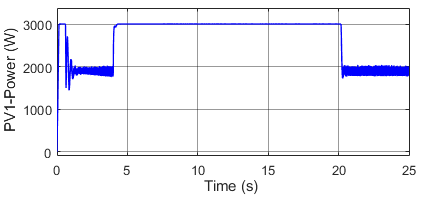} 
  \caption{PV-DG1 Power during Step Load.}
  \label{PV}
\end{subfigure}
\qquad
    \qquad
      \qquad
\begin{subfigure}{.4\textwidth}
  \centering
  % include second image
%\centering
%\includegraphics[scale=0.65]{pictures/DC link voltage2.png}
\includegraphics[width=\linewidth]{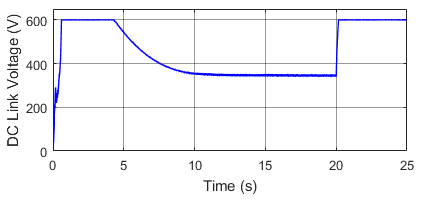}  
  \caption{DC link Voltage.}
  \label{DC}
\end{subfigure}
\caption{The performance of PV-DG1 power and DC link voltage.}
\label{PV-DC}
\vspace{-9pt}
\end{figure*}
\begin{figure*}[t]
\begin{minipage}{0.31\textwidth}
\includegraphics[width=\linewidth]{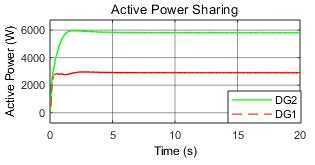}
\caption{\label{P}Sharing Performance\\ between PV-DGs for Active Power.}
\end{minipage}
\begin{minipage}{0.31\textwidth}
\includegraphics[width=\linewidth]{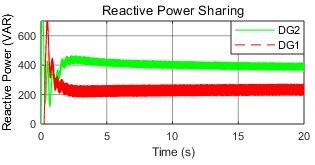}
\caption{\label{Q} Sharing Performance\\between PV-DGs for Reactive Power.} 
\end{minipage}
\begin{minipage}{0.35\textwidth}
\includegraphics[width=\linewidth]{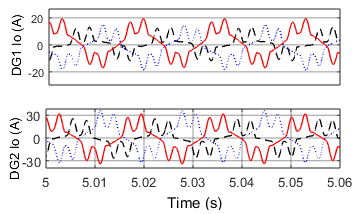}
\caption{\label{current} DG1 and DG2 current sharing.}
\end{minipage}
\vspace{-8pt}
\end{figure*}
Extracted components are received by the VCC in  Fig.~\ref{main} to reduce unbalanced voltage and harmonic distortion. The values of voltage unbalance factor ($VUF$) and positive and negative harmonic distortion indices ($HD^{h}$) are calculated as  
\begin{equation}
\begin{split}
\%VUF= \frac{\sqrt{(V_{dq}^{-1})^{2}}}{\sqrt{(V_{dq}^{+1})^{2}}}*100
\\
\%HD^{h}= \frac{(\sqrt{(V_{dq}^{h})^{2}}}{\sqrt{(V_{dq}^{+1})^{2}}}*100
\end{split}
\label{percent}
\end{equation}
where $V_{dq}^{-1}$, $V_{dq}^{+1}$ and $V_{dq}^{h}$ represent the magnitude of the negative and positive sequence of the fundamental voltage, and the main harmonic voltage in $dq$ frame at PCC respectively\cite{09,17}. 

$VUF_{ref}$ and $HD_{ref}^{h}$ in Fig.~\ref{main} are the reference of $VUF$ and  $h^{th}$ HD for the PCC voltage. The references are compared with $VUF$, $HD^{h}$. Note that if the unbalanced factor and harmonic references are less than PCC voltage distortion ($VUF$ and $HD^{h}$), saturation block must be used to not affect the stability of the control system. The errors are fed to a PI controller to reduce voltage unbalanced and harmonic distortion. Then each output of the PI controller of the negative sequence of the fundamental component and selected $h^{th}$ harmonic component is multiplied by $V_{dq}^{-1}$ and $V_{dq}^{h}$ respectively. Then, the signals are transformed from $dq$ frame to $\alpha \beta$ reference frame and added. The angular frequency generated by the active power controller is set to $-W_{ref}$ and $h*W_{ref}$ for the negative sequence of the fundamental component and selected $h^{th}$ harmonic component, respectively.
The references for voltage compensation ($V_{c}$) is generated as follows
\begin{equation}
\begin{split}
V_{c}=(((VUF_{ref}-VUF).PI.V_{dq}^{-1})+\\
\sum_{h=3,-5,+7,-11}((HD_{ref}^{h}-HD^{h}).PI.V_{dq}^{h})))\frac{P_{o,i}}{\sum_{k}^{n}P_{o,k}}
\end{split}
\label{Vc}
\end{equation}
where n is the number of DGs. The ratio of $DG_{i}$ rated power (${P_{o,i}}$) and the total power of the DG units (${\sum_{k}^{n}P_{o,k}}$) is needed to send the proper reference signal to all DG primary controllers considering their power capacities.
$V_{c}$ is transmitted to each primary local control to mitigate PCC distortion. In the control scheme, the VCC manages compensation of negative sequence of fundamental and main harmonic components.%(add table for PI)
\begin{figure}
\centering
\includegraphics[width=\linewidth]{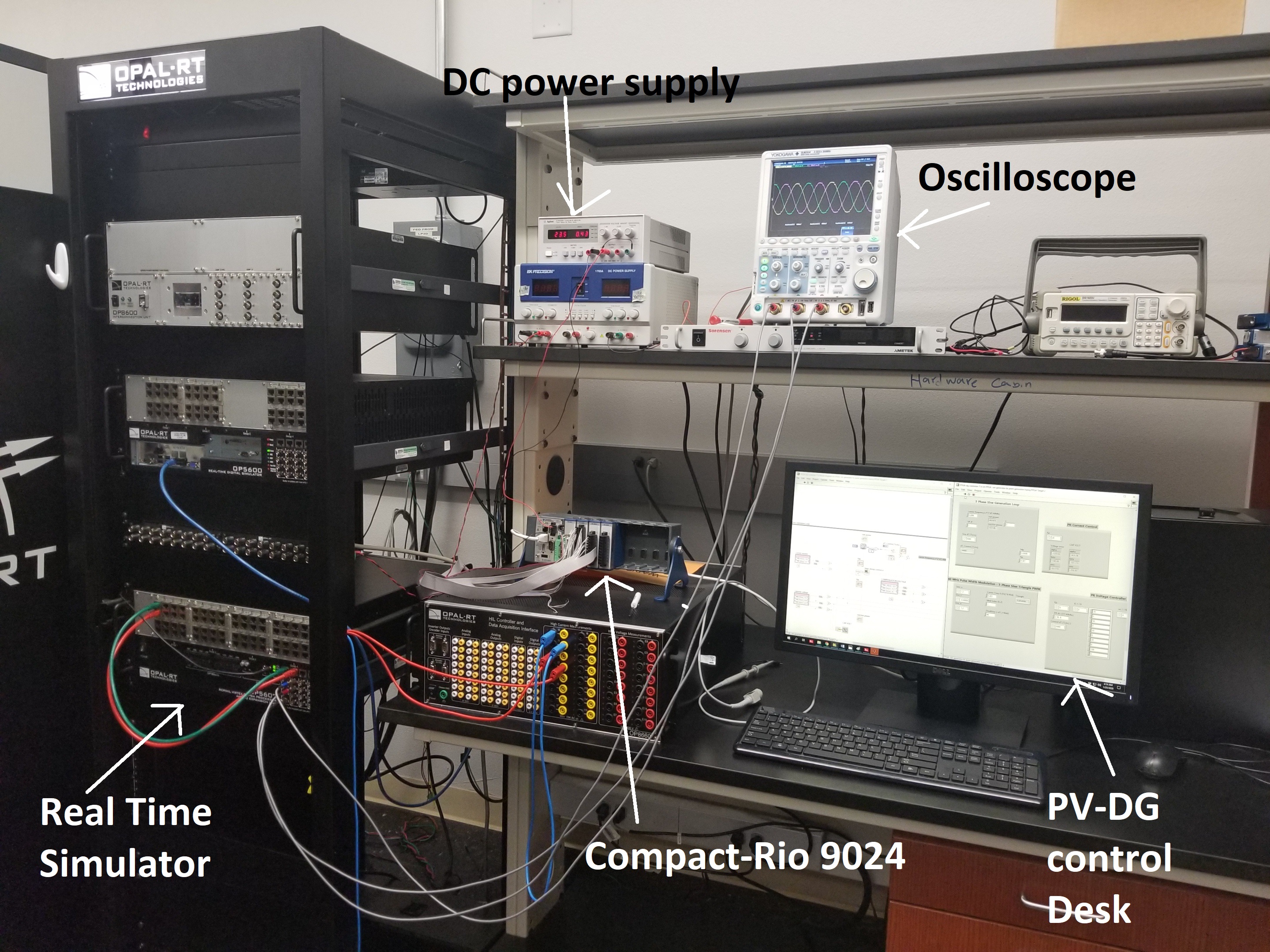}
\caption{Control Hardware in the loop in Real Time Simulation.}
\label{HIL}
\vspace{-10pt}
\end{figure}

\begin{figure*}[t]
\begin{subfigure}{.32\textwidth}
  \centering
  % include first image
 \includegraphics[width=\linewidth]{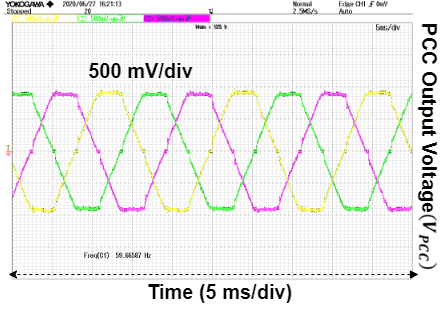} 
  \caption{}
\label{6a}
\end{subfigure}
\begin{subfigure}{.32\textwidth}
  \centering
  % include second image
  \includegraphics[width=\linewidth]{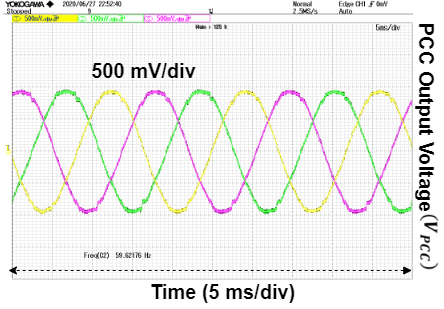}  
  \caption{}
\label{6b}
\end{subfigure}
\begin{subfigure}{.35\textwidth}
  \centering
  % include 3th image
  \includegraphics[width=\linewidth]{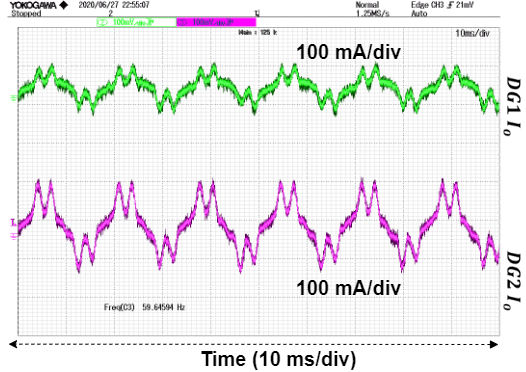}  
  \caption{DG1 and DG2 current sharing.}
  \label{7}
\end{subfigure}
\caption{Output Voltage of PCC in experimental study before (a) and after (b) compensation and (c) DG1 and DG2 current sharing.}
\label{R-Q}
\vspace{-16pt}
\end{figure*} 

\begin{figure}[ht]
\centering
\centering{\includegraphics[width=0.8\linewidth]{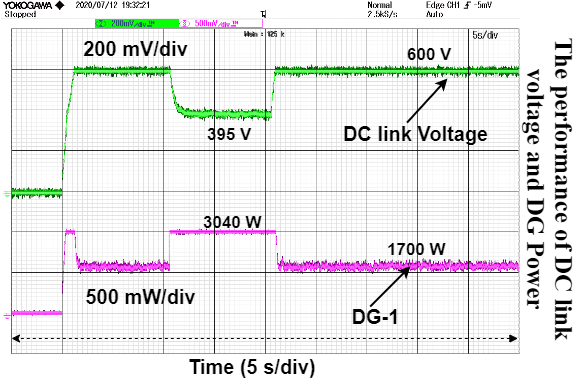}}
\caption{The performance of DC link voltage and DG Power during Step Load.}
\label{dc likn and pv power4}
\centering
\centering{\includegraphics[width=0.8\linewidth]{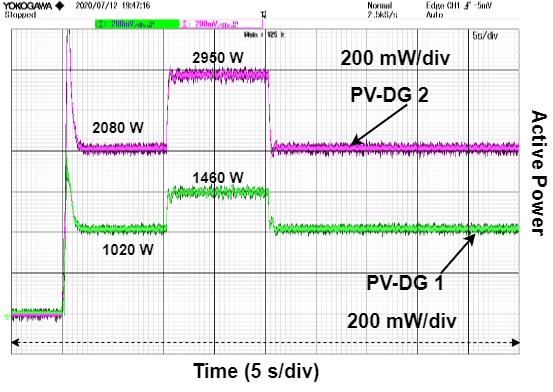}}
\caption{The performance of Power Sharing between DGs.}
\vspace{-15pt}
\label{power sharing3}
\end{figure}
%To summarize the control system in Fig. 2, the PCC voltage ($V_{pcc}$) unbalanced and harmonic data are extracted by the dq extraction block to send PCC voltage information to the VCC. In addition, The VCC transmits compensation references of voltage unbalance and harmonic components to all DG primary controller to improve the PCC voltage quality. Primary local controllers receive the compensation voltage signal from the VCC and produce the reference voltage for DGs interface inverters.
\section{Simulation and experimental results}
In order to verify the effectiveness of the proposed method, simulation and experimental results are presented using MATLAB/ Simulink and control hardware-in-the-loop (CHIL) respectively. The test system of the proposed islanded micro-grid shown in Fig.~\ref{main} consisting of the two parallel connected PV-DG units with unbalanced and nonlinear loads and connected to an IEEE 14 node distribution network is simulated. % which is acting as a microgrid. The DG2 unit is chosen to  be two times of DG1 for evaluating the performance of the control system.
DG2 unit is chosen to have double the power capacity of DG1. In addition, the power stage and control parameters are listed in Table~\ref{table}. The effectiveness of the proposed control scheme is tested under unbalanced and nonlinear loads.
%unbalanced voltage quality and reduction of harmonic voltage.
\subsection{Simulation Results}
This section presents the simulation results of the test system in Fig.~\ref{main} using MATLAB/ Simulink.
\subsubsection{Performance of VCC}
Fig.~\ref{Vpcc} and Fig.~\ref{thd} depict the PCC output voltage and the frequency spectrum before and after compensation respectively. Fig.~\ref{Vpcc}(a) shows the performance of the conventional micro-grid, with no the virtual impedance or VCC controller. The PCC output voltage is unbalanced and highly distorted before compensation. In addition, the harmonic distortion (HD) is high especially, the main harmonics ($3^{rd}$, $5^{th}$, $7^{th}$, and $11^{th}$) as depicted in Fig.~\ref{thd}(a). Voltage quality of the PCC is improved after compensation as shown in Fig.~\ref{Vpcc}(b). Furthermore, the total harmonic distortion (THD) is reduced as shown in Fig.~\ref{thd}(b) from 6.38\% to 1.91\%, which is less than 5\% maximum THD in compliance with the IEEE-519 standards \cite{10}. The voltage unbalanced factor ($VUF$) of PCC terminal was reduced from 5.8\% to 0.2\% with compensation. 
%when virtual impedance loops and VCC for unbalanced and nonlinear load compensation are not performed. 
%To activate the proposed method, the virtual impedance and the secondary control for unbalanced and nonlinear load compensation are adopted.
%Voltage quality of the PCC is improved effectively after compensation as shown in Fig.~\ref{Vpcc}(b). Furthermore, the reduction of harmonic distortion indices values for $3^{rd}$, $5^{th}$, $7^{th}$, and $11^{th}$  with the activation of the proposed VCC in Fig.~\ref{thd}(b) also indicates the total harmonic distortion ($THD$) being reduced from 6.38\% to 1.91\%, which is less than 5\% maximum THD in compliance with the IEEE-519 standards \cite{10}. The voltage unbalanced factor ($VUF$) of PCC terminal was reduced from 5.8\% to 0.2\% after the compensation. 
\subsubsection{Performance of VR controller} The performance of the VR controller was verified by applying a load step drop in the load at 4 seconds. It can be seen from Fig.~\ref{PV-DC} that PV power reached to maximum power between 4 and 20 seconds operating at the MPPT mode. Then when the load is dropped at 20 seconds, the VR controller became active and the DC link voltage was regulated to 600 V by curtailing the PV-DG power. It should be noted that, only 4\% of maximum output power of the PV array is curtailed in this case. 

\subsubsection{Performance of the load power sharing}  Active and reactive power shared proportionally between different rated PV-DGs are presented after compensation as shown in Fig.~\ref{P} and Fig.~\ref{Q} respectively and it should be noted that proper power sharing between DGs demonstrates the effectiveness of the droop controllers and the virtual impedance loop. Also output currents of DG1 and DG2 are presented in Fig.~\ref{current} to demonstrate the proper sharing of unbalanced and harmonic load current among the DGs after compensation. % This results are acceptable to supply two parallel connected DG units with the unbalanced plus nonlinear load.

\subsection{CHIL Experimental Validation}
This section demonstrates the results of the CHIL real-time simulation of the proposed system and controllers and % the test system similar to Fig.~\ref{main} is simulated with the control hardware in loop experiments 
using OPAL-RT as shown in Fig.~\ref{HIL}. The power stage including the DGs, inverters and loads are located in Mat-lab/Simulink. The proposed controllers are implemented using Compact Rio-9024 by National Instruments(NI) as the control hardware. The controller algorithm is coded in LabVIEW-FPGA. %as shown in Fig.~\ref{HIL}.
The experimental results of the CHIL simulation are shown in Fig.~\ref{R-Q}-\ref{power sharing3}. It can be observed that the results of the experimental study also demonstrate a noticeable improvement in the power quality and sharing. Furthermore, the system is stable under load step change. 
\section{conclusion}
A novel voltage source inverter controller is proposed for PV islanded micro-grid under unbalanced and non-linear loads. The proposed controller performance is verified with simulation as the PCC voltage distortion is decreased by 70\%, while the load power-sharing is achieved among the PV inverters. In addition, fixing the DC link voltage is achieved by the DC voltage regulator controller. The effectiveness of the proposed control scheme is validated using Opal-RT real-time simulation of an IEEE-14 bus distribution system and experimentally using control hardware-in-the-loop.
%A novel voltage controlled inverter for combined operation of the voltage compensation controller and the DC voltage regulator controllers is proposed in the PV islanded micro-grid. It has been shown that the voltage controlled compensation method can operate under unbalanced nonlinear loads in the islanded PV micro-grids.  Fixing the DC link voltage is achieved by the DC voltage regulator controller. Moreover, simulation results verified the effectiveness of the proposed control method as the PCC voltage distortion is decreased by 70\%, while the load power-sharing is achieved among the PV inverters. The effectiveness of the proposed control scheme is validated using Opal-RT real-time simulation of an IEEE-14 bus distribution system and experimentally using control hardware-in-the-loop. The complete simulation and experimental results are discussed.
\section*{Acknowledgment}
The authors would like to thank Mohammad Khatibi and Reynaldo Gonzalez for their assistance and useful suggestions throughout this paper.
\bibliographystyle{IEEEtran}
\bibliography{IEEEexample}
\end{document}